\documentclass[useAMS,usenatbib]{mn2e}
\usepackage{eso-pic,graphicx}
\newcommand{\bjdtdb}{\ensuremath{\rm {BJD_{TDB}}}}

\newcommand{\mj}{\ensuremath{\,M_{\rm J}}}
\newcommand{\rj}{\ensuremath{\,R_{\rm J}}}
\newcommand{\fave}{\langle F \rangle}
\newcommand{\fluxcgs}{10$^9$ erg s$^{-1}$ cm$^{-2}$}

\title[Characterization of WASP-5b, WASP-44b and WASP-46b]
      {Multi-band characterization of the hot Jupiters: WASP-5b, WASP-44b, and WASP-46b}

\author[M. Moyano et al.]
       {M.~Moyano$^{1}$\thanks{E-mail: mmoyano@ucn.cl},
         L.~A.~Almeida$^{2}$\thanks{E-mail: leonardodealmeida.andrade@gmail.com},
         C. von Essen$^{3}$, F.~Jablonski$^{4}$, M.~G.~Pereira$^{5}$ \\
         $^1$Instituto de Astronom\'ia, Universidad Cat\'olica del Norte, 
         Av. Angamos 0610, Antofagasta, Chile.\\
         $^2$Instituto de Astronomia, Geof\'isica e Ci\^encias Atmosf\'ericas, Universidade de S\~ao Paulo, \\
         Rua do Mat\~ao 1226, Cidade Universit\'aria S\~ao Paulo-SP, 05508-090, Brasil \\
         $^3$Stellar Astrophysics Centre, Department of Physics and Astronomy, Aarhus University,
         Ny Munkegade 120, 8000, Aarhus C, Denmark\\
         $^4$Instituto Nacional de Pesquisas Espaciais/MCTI, Avenida dos Astronautas 1758,
         S\~ao Jos\'e dos Campos, SP, 12227-010, Brasil\\
         $^5$Departamento de F\'isica, Universidade Estadual de Feira de Santana, 
         Av. Transnordestina, S/N, Feira de Santana, BA, 44036-900, Brasil \\
       }
       
\date{Accepted XXXX. Received YYYYYY; in original form 2017 May}
   \pagerange{\pageref{firstpage}--\pageref{lastpage}} \pubyear{2017}
\begin{document}
\label{firstpage}
\maketitle
\begin{abstract}
  We have carried out a campaign to characterize the hot Jupiters WASP-5b, WASP-44b, and WASP-46b using
  multi-band photometry collected at the Observat\'orio do Pico Dos Dias in Brazil. We have determined
  the planetary physical properties and new transit ephemerides for these systems. The new orbital parameters
  and physical properties of WASP-5b and WASP-44b are consistent with previous estimates. In the case of
  WASP-46b, there is some quota of disagreement between previous results. We provide a new determination
  of the radius of this planet and help clarify the previous differences. We also studied the transit time 
  variations including our new measurements. No clear variation from a linear trend was found for 
  the systems WASP-5b and WASP-44b. In the case of WASP-46b, we found evidence of deviations indicating the 
  presence of a companion but statistical analysis of the existing times points to a signal due to the sampling 
  rather than a new planet. Finally, we studied the fractional radius variation as a function of wavelength 
  for these systems. The broadband spectrums of WASP-5b and WASP-44b are mostly flat. In the case of WASP-46b
  we found a trend, but further measurements are necessary to confirm this finding.

\end{abstract}

\section{INTRODUCTION}
Since the announcement of the discovery of the first exoplanets orbiting the pulsar PSR~1257+12
\citep{WolszczanFrail1992,Wolszczan1994} and the subsequent discovery of an exoplanet around a
main sequence star \citep{Mayor1995}, more than three thousand planets orbiting other stars have
been detected \citep[see exoplanet.eu]{Schneider2011}. An exoplanet's orbit oriented along the line
of sight provides unmatchable access to a list of both planetary astrophysical properties and orbital
elements. The favorable geometry of transiting extrasolar planets (TEPs) allows the measurement of
the true planetary  masses and radii provided some external constraints from stellar evolutionary
theory or empirical stellar mass-radius relations \citep{Southworth2009}. Moreover, if the stellar
limb-darkening is assumed to be negligible, the density of the star can be measured directly from the
light curve alone \citep{Seager2003}. Additionally, some of the stellar light filters through the
exoplanet atmosphere during transit, resulting in spectral footprints that may reveal characteristics
of the exoplanetary atmosphere \citep[e.g.][]{Swain2010}. TEPs allow for the first time to access an
accurate ensemble of planetary properties, thus the accurate characterization of each system plays
an important role for both the determination of fundamental parameters and recognition of
astrophysically interesting targets for follow-up work such as Transit Timing Variation
\citep[TTV]{Holman2005,Holman2010} or atmospheric characterization (see e.g., \citealt[]{Seager2010ARAA}).\\
Hot Jupiters (HJs) are a class of large and gaseous planets similar to Jupiter, but orbiting very close
to their host stars. These characteristics make these objects easier to detect compared with low-mass
planets at wider orbits. Despite the large number of HJs discovered fundamental questions about their
formation and evolution are still under debate. The main scenarios proposed to explain their
tight orbital trajectories are the formation of orbits with a small inclination via disk migration, or
very inclined orbits via high-eccentricity migration due to tidal dissipation caused by gravitational
interactions with another companion \citep[see e.g.][]{FordRasio2008,Thies2011,TutukovFedorova2012,
ValsecchiRasio2014}. Therefore, precise physical and geometrical parameters of HJs derived from optical and 
infrared photometry are essential to test the planetary formation and evolution theories as well as 
to distinguish different types of atmospheres \citep[][]{Sing2016}. \\
In this paper we present precision relative photometry of three HJs: WASP-5b, WASP-44b and,
WASP-46b \citep{Anderson2008,Anderson2012}. WASP-5b is a Hot-Jupiter with a mass of $M_P = 1.64 \mj$ and
a radius of $R_P = 1.17 \rj$ transiting a bright (V = 12.3 mag) G4V star on a $1.63$ day orbit. WASP-44b
has a mass of $M_P = 0.89 \mj$ and a radius of $R_P = 1.00 \rj$ orbiting a bright (V = 12.9 mag) G8V star
with an orbital period of $2.42$ day. WASP-46b is a massive ($M_P = 2.10 \mj$) hot Jupiter with a radius
of  $R_P = 1.31 \rj$ eclipsing a bright (V = 12.9 mag) G6V star on a $1.43$ day orbit.\\
This paper is structured as follows: Section 2 describes the observations and the data reduction,
Section 3 presents our results and analysis, and in Section 4 we discuss our results and state
our conclusions.
\section{OBSERVATIONS AND DATA REDUCTION}
The observations were carried out using the facilities of the Observat\'orio do Pico dos Dias (OPD/LNA), 
in Brazil\footnote{http://www.lna.br/opd/opd.html}. Transits of the planets WASP-5b, WASP-44b, 
and WASP-46b were observed on 2011 August, 2012 August, 2013 July, and 2013 August using
the Andor iKon-L CCD cameras mounted on the 1.6-m and 0.6-m telescopes. These cameras provide plate
scales of 0.18 and 0.34 arcsec/pixel, respectively. A summary of the collected data is reported in
Table~\ref{LOG_observations}. In this table, $N$ is the number of individual images obtained with
integration time t$_{\rm exp}$.\\
\begin{table*}
\centering
\begin{minipage}{0.9\textwidth}
\caption{Summary of the photometric observations presented in this work.}
\label{LOG_observations}
\begin{tabular}{l l c c c c c c c}
\hline\hline
Target   & UT Date      & N & t$_{\rm exp}$(s) & Telescope & Filter & Aperture(pix) &
Scatter(\%)\footnote{Standard deviation of the residuals after subtracting the fitted model} &
Slope Spectrum\footnote{See section\ref{broadband_spectrum}}\\
\hline
\hline
 WASP-5b  & 2012 Aug 10  & 992 & 10      & 1.6-m    & V            & 16.0  & 0.23 & $(1.73 \pm 2.3)\times 10^{-5}$\\
          & 2012 Aug 10  & 362 & 40      & 0.6-m    & I$_{\rm C}$   & 9.0   & 0.26 &\\
\hline
WASP-44b  & 2012 Aug 11  & 296 & 60      & 1.6-m    & V            & 11.0  & 0.14 &\\
          & 2012 Aug 11  & 346 & 40      & 0.6-m    & R$_{\rm C}$   & 8.0   & 0.38 &\\
          & 2012 Aug 11  & 200 & 60      & 0.6-m    & I$_{\rm C}$   & 5.0   & 0.40 & $(-1.07 \pm 11.1)\times 10^{-6}$\\
          & 2012 Aug 16  & 230 & 60      & 1.6-m    & B            & 11.0  & 0.19 &\\
          & 2013 Aug 01  & 832 & 08      & 1.6-m    & I$_{\rm C}$   & 7.0   & 0.24 &\\
\hline
WASP-46b  & 2011 Aug 14  & 300 & 30      & 1.6-m    & V            & 13.0  & 0.19 &\\
          & 2013 Jul 30  & 601 & 10      & 1.6-m    & I$_{\rm C}$   & 10.0  & 0.23 & $(-2.31 \pm 1.77)\times 10^{-5} $\\
          & 2013 Aug 02  & 361 & 15      & 1.6-m    & R$_{\rm C}$   & 14.0  & 0.17 &\\ 
\hline
\end{tabular}
\end{minipage}
\end{table*}
The basic data reduction was done using {\tt IRAF}\footnote{http://www.iraf.noao.edu} tasks. 
We created a master median bias of typically $\sim$100 bias frames for each observing night.
A normalized master flat-field frame was obtained by combining and then normalizing $\sim$30 dome
flat-field images. We process the images by subtracting the master bias and then dividing by
the normalized master flat frame.\\
The target fields are not crowded, thus we performed standard differential aperture photometry.
The telescope tracking and pointing were not stable during the observations, thus we carefully
placed the apertures in each image for both the target and the reference non-variable stars.
The fluxes were extracted using an implementation of DAOPHOT \citep{Stetson1987}. We experimented
with different apertures and sky rings and kept those which resulted in the lowest standard deviation 
after subtracting the fitted transit model (see next section). This was an efficient way of both
measuring the fit quality, due to the reduced number of out-of-transit observations, and discarding outliers.
We iteratively removed the outliers discarding measurements $3 \times \sigma$ away from the resulting
model light curve (see next section). We repeated this process up to 3 times for each band which resulted
in a maximum of removal of 1.5\% of the data points and an average reduction of 10\% in the residuals standard
deviation. The final aperture values used to extract our light curves are listed in Table~\ref{LOG_observations}
\footnote{The original light curves can be downloaded at http://www.iaucn.cl/Users/Max/PUBLIC/PAPER\_OPD/}.
\section{Analysis and Results} 
\subsection{Light curve Analysis}
\label{light_curve_analysis}
\begin{figure}
\resizebox{\hsize}{!}{\includegraphics[angle=270,bb=70 140 550 590]{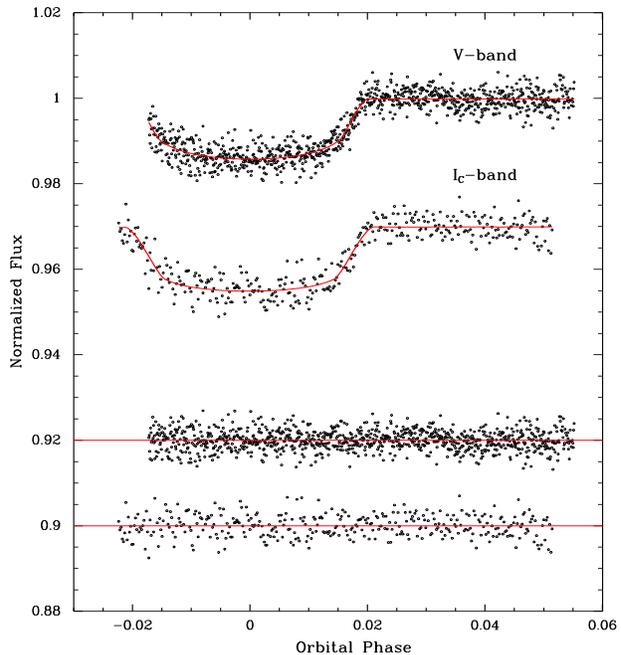}} \\
\caption{WASP-5b light curves. From top to bottom the V- and I$_C$-band
  light curves respectively. The red curves are the best fits superimposed. The residuals of
  the fitted model are displayed at the bottom. The displayed light curves are decorrelated using
the parameters described in Section \ref{light_curve_analysis}.}
\label{wasp-5b_lc}
\end{figure}

\begin{figure}
\resizebox{\hsize}{!}{\includegraphics[angle=270,bb=70 140 550 590]{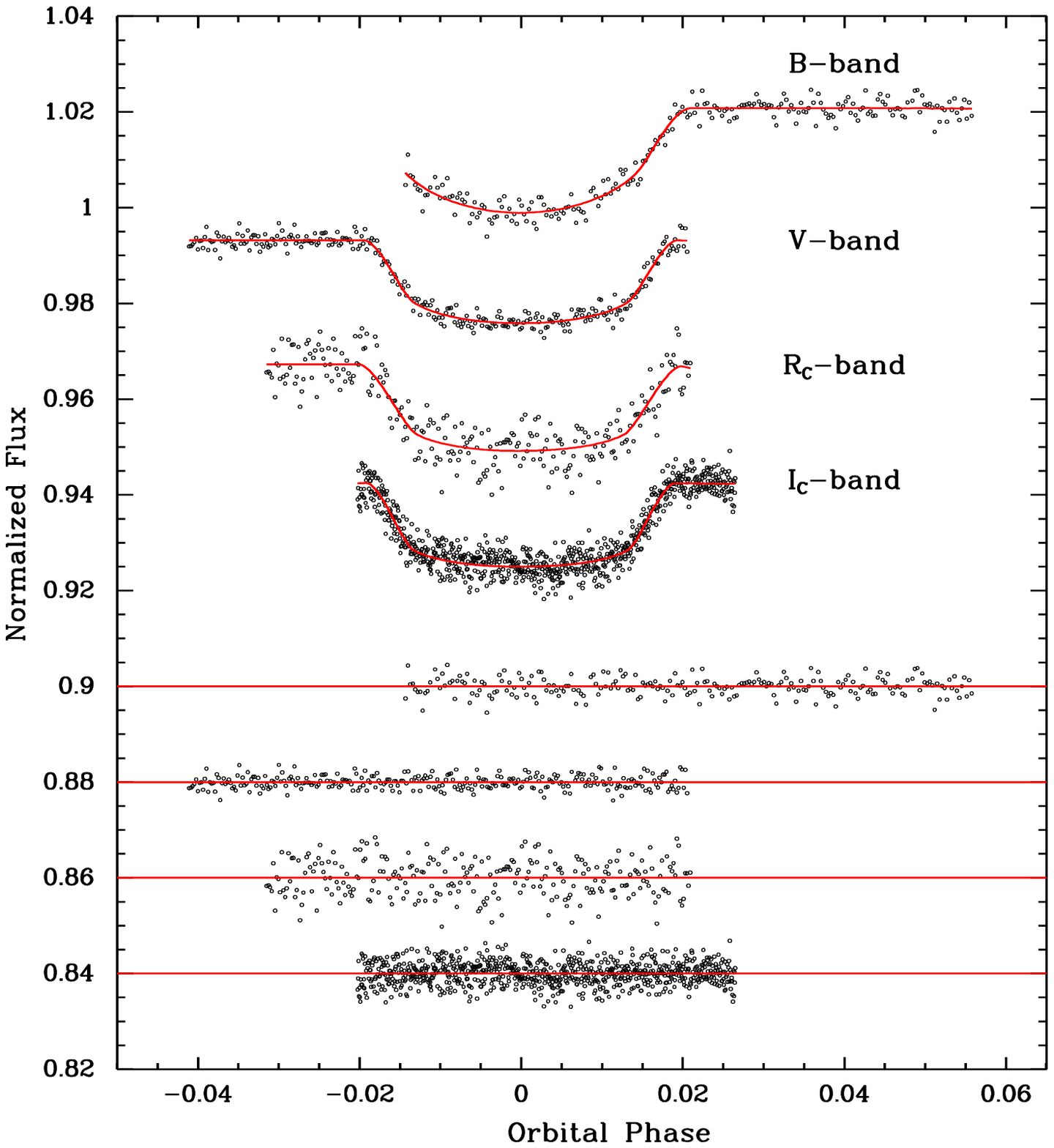}} \\
 \caption{WASP-44b light curves. From top to bottom the B-, V-, R$_C$-, and I$_C$-band light
curves respectively. The red curves are the best fits superimposed. The residuals of the
fitted model are displayed at the bottom. The displayed light curves are decorrelated using
the parameters described in Section \ref{light_curve_analysis}. }
\label{wasp-44b_lc}
\end{figure}

\begin{figure}
 \resizebox{\hsize}{!}{\includegraphics[angle=270,bb=70 140 550 590]{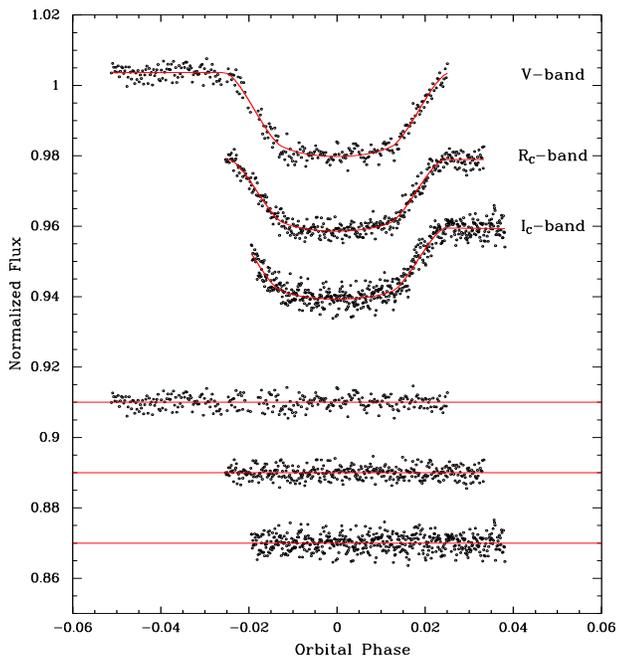}}\\
 \caption{WASP-46b light curves. From top to bottom the V-, R$_C$, and I$_C$-band
light curves respectively. The red curves are the best fits superimposed. 
The residuals of the fitted model are displayed at the bottom. The displayed light curves are
decorrelated using the parameters described in Section \ref{light_curve_analysis}.}
\label{wasp-46b_lc}
\end{figure}

To fit the light curves we used EXOFAST \citep{Eastman2013}. This software implements the Markov Chain Monte
Carlo (MCMC) method to estimate and characterize the parameters' uncertainty distributions \citep{Ford2005,Ford2006}
and the light curve models of \cite{Mandel2002}. To evolve the MCMC chains it uses the Differential Evolution MCMC
method \citep{terBraak2006}. We included at each MCMC step the background value at the central pixel, airmass,
and CCD positions, thus ensuring these systematics are considered at each MCMC jump. In fact, these systematics
were critical to fit properly our light curves in the cases of changes of the positions of our targets and references
(see discussion below about our systematics). To include these parameters in the MCMC chains we considered
a linear combination of the (x,y) pixel positions of the target's star centroid, the airmass, and the
background value which was computed averaging the pixel values within a ring around the star. For each light
curve we ran MCMC chains of maximum 100,000 iterations, and discarded the first 25,000 chains which eliminated any bias
due to the starting conditions (``burn-in'' process, \citealt{Tegmark2004}). To check their convergency, we divided
the final 75,000 values into three, and computed the mean and standard deviation (best fit value and error) of
the posterior distributions. If the derived values were consistent within 1 sigma errors, we considered the chains
to be convergent. Since we fitted only transit data, we complemented our fits with spectroscopic information taken
from \citet{Anderson2008,Anderson2012,Triaud2010}. We fixed the eccentricities to zero, as the radial velocity data
is consistent with circular orbits.\\
EXOFAST uses a quadratic law to take into account the limb-darkening. \citet{Southworth2008} proved this law
is sufficient to model high quality ground-based light curves. We fitted both coefficients for each light curve.
\citet{Espinoza2015} proved that this approach introduces bias just up to $\sim 1\%$ in $R_p/R_*$. EXOFAST
adds a penalty term at each MCMC jump based on the tables of \citet{Claret2011}, thus constraining the fitted
limb-darkening coefficients and preventing their values to be unphysical.\\
We calculated the standard deviation of the measurements after subtracting the fitted model (scatter) to
assess the quality of the fit. The final scatter values are displayed on Table~\ref{LOG_observations}.\\
Light curves of planetary transits have systematic effects, the so called correlated ``red noise''
\citep{Pont2006}. To quantify this noise we used the residual permutation method \citep[RPM]{Jenkins2002}
implemented in the transit analysis package JKTEBOP \citep{Southworth2008}. In this method, the residuals
around the best fit are shifted point by point along the observations and a new fit is performed (if a
residual is shifted after the end it is moved to the beginning). This is done for all observational
data points. From the resultant distribution the errors are calculated as in the MCMC case.
These uncertainties are a way to quantify the systematics present in our data. We found that the
uncertainties determined by the RPM are in most cases similar to the MCMC case for
the data taken with the $1.6-m$ telescope (there just a couple of extreme cases where the RPM uncertainties
are slightly larger). In the case of data taken with the $0.6-m$ telescope, the RPM uncertainties are up to
three times the MCMC uncertainties, so the systematics are stronger in this case.
This is in part explained because the pointing of the $0.6-m$ telescope changes more drastically (drift
of 90 pixels for every 20 minutes of observing time), thus target and references are placed on different
CCD positions, therefore increasing the systematics due to changes in quantum efficiency. Additionally,
the $0.6-m$ has no guiding system. The $1.6-m$ telescope has a guiding system and therefore a better pointing.
As pointed out before, these changes of position were taken into account at each MCMC iteration.\\
We also fitted the light curves with JKTEBOP using MCMC simulations (using an approach similar to that of
EXOFAST). This served as a consistency  check of our EXOFAST results. We found the results to be consistent
with each other within one standard deviation. We preferred EXOFAST results because it implements at each MCMC
jump a penalty term taking into account the decorrelation parameters. As stated before, taking into account
the systematics present in our data is critical for improving the final results and to properly characterize
the uncertainties. As a final consistency check, we also analyzed our light curves using the graphical transit analysis
interface TAP \citep{Gazak2012}. We also found the results to be consistent with EXOFAST within one
standard deviation. This is expected, because TAP uses EXOFAST \citep{Gazak2012}, but iterates the MCMC
chains using a wavelet-based likelihood \citep{Carter2009}. All the results presented in this paper
come from the EXOFAST results and the other tools were used just as an initial consistency check.
Figures~\ref{wasp-5b_lc}, \ref{wasp-44b_lc} and \ref{wasp-46b_lc} show the resultant light curves of
WASP-5b, WASP-44b, and WASP-46b, respectively. In all cases, the scatter of the residuals was better
than 0.4\%.\\
\subsection{System Parameters}
\label{system_parameters}
Tables~\ref{wasp5b_table_results}, ~\ref{wasp44b_table_results}, and ~\ref{wasp46b_table_results} show
the determined parameters in all bands for the planets WASP-5b, WASP-44b, and \mbox{WASP-46b}, respectively.
As mentioned before, the quality of the light curves taken with the 1.6-m telescope is better, so we
preferred those for I$_C$-band results of WASP-44b. Thus, all the results presented in this
paper refer to the I$_C$-band taken with the 1.6-m telescope for the WASP-44b system.
\begin{table*}
\centering
\begin{minipage}{0.7\textwidth}
\caption{Planetary parameters of WASP-5b.}
\label{wasp5b_table_results}
\begin{tabular}{ l l  l l }
\hline
Symbol & Parameter & V & I$_C$\\
\hline
                       ~~~$a$\dotfill  & Semi-major axis (AU)\dotfill  &  $0.02789_{-0.00058}^{+0.00059}$ &  $0.02793_{-0.00055}^{+0.00056}$\\
                           ~~~$R_{P}$\dotfill  & Radius (\rj)\dotfill  &  $1.126_{-0.082}^{+0.100}$ &  $1.240_{-0.081}^{+0.085}$\\
                        $R_{P}/R_{\star}$\dotfill  & Planet/star radius ratio\dotfill &  $0.1114_{-0.0015}^{+0.0015}$ &  $0.1158_{-0.0017}^{+0.0018}$\\
           ~~~$T_{eq}$\dotfill  & Equilibrium Temperature (K)\dotfill  &  $1753_{-62}^{+70}$ &  $1765_{-62}^{+64}$\\
               ~~~$\fave$\dotfill  & Incident flux (\fluxcgs)\dotfill  &  $2.14_{-0.29}^{+0.37}$ &  $2.21_{-0.29}^{+0.34}$\\
                ~~~$T_C$\dotfill  & Time of mid-transit(\bjdtdb-2450000)\dotfill  &  $6150.61479_{-0.00056}^{+0.00050}$ &  $6150.61396_{-0.00057}^{+0.00054}$\\
              ~~~$u_1$\dotfill  & linear limb-darkening coeff\dotfill  &  $0.484\pm0.049$ &  $0.285_{-0.050}^{+0.048}$\\
           ~~~$u_2$\dotfill  & quadratic limb-darkening coeff\dotfill  &  $0.283\pm0.052$ &  $0.290_{-0.049}^{+0.048}$\\
                      ~~~$i$\dotfill  & Inclination (degrees)\dotfill  &  $86.1_{-1.4}^{+1.8}$ &  $84.54_{-0.66}^{+0.74}$\\
                         ~~~$\delta$\dotfill  & Transit depth\dotfill  &  $0.01155_{-0.00051}^{+0.00056}$ &  $0.01373_{-0.00052}^{+0.00053}$\\
                 ~~~$T_{14}$\dotfill  & Total duration (days)\dotfill  &  $0.0978_{-0.0016}^{+0.0018}$ &  $0.0935_{-0.0023}^{+0.0025}$\\
\hline
\end{tabular}
\end{minipage}
\end{table*}
\begin{table*}
\centering
\begin{minipage}{\textwidth}
\caption{Planetary parameters of WASP-44b.}
\label{wasp44b_table_results}
\begin{tabular}{ l l  l l l l }
\hline
Symbol & Parameter & V & B & R$_C$ & I$_C$\\
\hline
                       ~~~$a$\dotfill  & Semi-major axis (AU)\dotfill  &  $0.0352\pm0.0014$ &  $0.0328_{-0.0015}^{+0.0016}$ &  $0.0354\pm0.0016$ &  $0.0349_{-0.0016}^{+0.0015}$\\
                           ~~~$R_{P}$\dotfill  & Radius (\rj)\dotfill  &  $1.12_{-0.11}^{+0.10}$ &  $1.08_{-0.13}^{+0.15}$ &  $1.18\pm0.13$ &  $1.121_{-0.081}^{+0.080}$\\
                                                      $R_{P}/R_{\star}$\dotfill  & Planet/star radius ratio\dotfill &  $0.1224_{-0.0013}^{+0.0013}$ &  $0.1346_{-0.0031}^{+0.0031}$ &  $0.1256_{-0.0032}^{+0.0033}$ & $0.1246_{-0.0025}^{+0.0025}$ \\
           ~~~$T_{eq}$\dotfill  & Equilibrium Temperature (K)\dotfill  &  $1390\pm120$ &  $1200_{-100}^{+120}$ &  $1410_{-130}^{+140}$ &  $1360\pm120$\\
               ~~~$\fave$\dotfill  & Incident flux (\fluxcgs)\dotfill  &  $0.85_{-0.25}^{+0.33}$ &  $0.46_{-0.14}^{+0.23}$ &  $0.89_{-0.29}^{+0.40}$ &  $0.78_{-0.23}^{+0.31}$\\
                ~~~$T_C$\dotfill  & Time of mid-transit(\bjdtdb-2450000)\dotfill  &  $6151.82559_{-0.00045}^{+0.00046}$ &  $6156.6694_{-0.0019}^{+0.0014}$ &  $6151.82415_{-0.00095}^{+0.0010}$  &  $6505.70010\pm0.00025$\\
              ~~~$u_1$\dotfill  & linear limb-darkening coeff\dotfill  &  $0.493_{-0.084}^{+0.094}$ &  $0.932_{-0.12}^{+0.089}$ &  $0.378_{-0.089}^{+0.10}$ &  $0.331_{-0.070}^{+0.071}$\\
           ~~~$u_2$\dotfill  & quadratic limb-darkening coeff\dotfill  &  $0.222_{-0.088}^{+0.076}$ &  $-0.044_{-0.095}^{+0.12}$ &  $0.252_{-0.074}^{+0.066}$ &  $0.252_{-0.063}^{+0.061}$\\
                      ~~~$i$\dotfill  & Inclination (degrees)\dotfill  &  $86.13_{-0.58}^{+0.81}$ &  $86.89_{-0.85}^{+1.2}$ &  $85.85_{-0.60}^{+0.74}$ &  $86.23_{-0.48}^{+0.56}$\\
                      ~~~$\delta$\dotfill  & Transit depth\dotfill  &  $0.01501_{-0.00079}^{+0.00065}$ &  $0.0171_{-0.0014}^{+0.0015}$ &  $0.0163\pm0.0012$ &  $0.01542_{-0.00070}^{+0.00071}$\\
                 ~~~$T_{14}$\dotfill  & Total duration (days)\dotfill  &  $0.0942\pm0.0016$ &  $0.0963_{-0.0038}^{+0.0042}$ &  $0.0937_{-0.0030}^{+0.0032}$ &  $0.09516_{-0.00095}^{+0.00099}$\\
\hline
\end{tabular}
\end{minipage}
\end{table*}
\begin{table*}
\centering
\begin{minipage}{\textwidth}
\caption{Planetary parameters of WASP-46b.}
\label{wasp46b_table_results}
\begin{tabular}{ l l  l l l }
\hline
Symbol & Parameter & V & R$_C$ & I$_C$\\
\hline
                       ~~~$a$\dotfill  & Semi-major axis (AU)\dotfill  &  $0.02421\pm0.00052$ &  $0.02414_{-0.00054}^{+0.00055}$ &  $0.02407\pm0.00055$\\
                           ~~~$R_{P}$\dotfill  & Radius (\rj)\dotfill  &  $1.338_{-0.062}^{+0.064}$ &  $1.230_{-0.051}^{+0.052}$ &  $1.209_{-0.072}^{+0.074}$\\
                           $R_{P}/R_{\star}$\dotfill  & Planet/star radius ratio\dotfill &  $0.1507_{-0.0017}^{+0.0017}$ &  $0.14109_{-0.00099}^{+0.00099}$ &  $0.1403_{-0.0022}^{+0.0021}$\\
           ~~~$T_{eq}$\dotfill  & Equilibrium Temperature (K)\dotfill  &  $1678_{-53}^{+51}$ &  $1657_{-52}^{+53}$ &  $1641\pm58$\\
               ~~~$\fave$\dotfill  & Incident flux (\fluxcgs)\dotfill  &  $1.80_{-0.22}^{+0.23}$ &  $1.71_{-0.20}^{+0.23}$ &  $1.65_{-0.22}^{+0.25}$\\
                ~~~$T_C$\dotfill  & Time of mid-transit (\bjdtdb-2450000)\dotfill  &  $5788.52807_{-0.00030}^{+0.00032}$ &  $6506.57629_{-0.00025}^{+0.00023}$ &  $6503.71529\pm0.00034$\\
              ~~~$u_1$\dotfill  & linear limb-darkening coeff\dotfill  &  $0.384_{-0.053}^{+0.055}$ &  $0.345_{-0.054}^{+0.055}$ &  $0.307_{-0.053}^{+0.054}$\\
           ~~~$u_2$\dotfill  & quadratic limb-darkening coeff\dotfill  &  $0.241_{-0.054}^{+0.053}$ &  $0.275\pm0.051$ &  $0.287_{-0.052}^{+0.049}$\\
                      ~~~$i$\dotfill  & Inclination (degrees)\dotfill  &  $82.87_{-0.32}^{+0.34}$ &  $82.73_{-0.32}^{+0.33}$ &  $82.92_{-0.41}^{+0.46}$\\
                      ~~~$\delta$\dotfill  & Transit depth\dotfill  &  $0.02293_{-0.00076}^{+0.00082}$ &  $0.01976_{-0.00043}^{+0.00045}$ &  $0.01943_{-0.00095}^{+0.0010}$\\
                 ~~~$T_{14}$\dotfill  & Total duration (days)\dotfill  &  $0.0727_{-0.0013}^{+0.0014}$ &  $0.06994_{-0.00094}^{+0.0010}$ &  $0.0703\pm0.0017$\\
\hline
\end{tabular}
\end{minipage}
\end{table*}
The inclination of the system and the orbital semi-major axis are independent of wavelength, thus a unique
value can be determined. We calculated a final value using as weight the inverse square of the individual
uncertainties. Table~\ref{weighted_mean_parameters} shows the results and previous estimates for
the planets WASP-5b, WASP-44b, and WASP-46b, respectively. Our values are consistent with previous results
for these planets. We have improved the value of the inclination for WASP-5b and the normalized
semi-major axis $a/R_*$ for WASP-46b.

\begin{table*}
\centering
\begin{minipage}{0.8\textwidth}
\caption{Weighted mean of common parameters for WASP-5b, WASP-44b, and WASP-46b}
\label{weighted_mean_parameters}
\begin{tabular}{l c c c c}
\hline
WASP-5b\\
\hline
Parameter  &  This work        & \citet{Southworth2009}   & \citet{Triaud2010} & \citet{Fukui2011}\\
\hline
i[degrees] &  $84.77\pm0.68$   & $85.8\pm1.1$  & 86.2$^{+0.8}_{-1.7}$   & 85.58$^{+0.81}_{-0.76}$    \\
$a/R_{*}$  &  $5.54\pm0.19$    & $5.41^{+0.17}_{-0.18}$ & 5.49$^{+0.37}_{-0.12}$   & 5.37$\pm$0.15  \\
\hline
WASP-44b\\
\hline
Parameter  &  This work        & \citet{Anderson2012}    & \citet{Mancini2013} & \citet{Turner2016}\\  
\hline
i[degrees] &  $86.18\pm0.37$  & 86.02$^{+1.11}_{-0.86}$ & $86.59\pm0.18$  &  \\
$a/R_{*}$  &   $8.10\pm0.20$   & 8.05$^{+0.66}_{-0.52}$  & $8.58\pm0.3 $   & $8.33^{+0.09}_{-0.14}$ \\
\hline
WASP-46b\\
\hline
Parameter  &  This work      & \citet{Anderson2012}  & \citet{Ciceri+2016}   \\           
\hline
i[degrees] &  $82.82\pm0.21$ &   $82.63\pm0.38$      &  $82.80 \pm 0.17$     \\
$a/R_{*}$  &  $5.76\pm0.09$  &   $5.74\pm0.15$       &   $5.85 \pm 0.12$\footnote{Value calculated using the
reported values of \citet{Ciceri+2016} (propagating the lowest reported uncertainties for $a$ and $R_{*}$)}\\
\hline
\end{tabular}
\end{minipage}
\end{table*}
\subsection{Transit ephemerides}
\label{Transit_ephemerides}
We determined the transit ephemerides of WASP-5b, WASP-44b, and WASP-46b assuming linear ephemerides.
We combined our mid-transit times with all measurements published in the literature including the times
available on TRESCA\footnote{The TRansiting ExoplanetS and CAndidates (TRESCA) web-site can be found
  at http://var2.astro.cz/EN/tresca/index.php}. We selected all the entries on TRESCA with a data
quality index better than 3 and we inspected each light curve for possible systematics. We converted all
measurements to barycentric dynamical time BJD(TDB) using the time routines from \cite{Eastman2010}.
For the TRESCA transit times we convert their JD(UTC) to BJD(TDB) which is more precise than converting
directly their HJD(TT) to BJD(TDB). To fit the mid-transit times we used the expression
$T_{\rm min} = T_0 + E \times P_{\rm b}$, where $T_{\rm min}$ are the predicted mid-transit times,
$T_0$ is a fiducial epoch, $E$ is the cycle count from $T_0$, and $P_{\rm b}$ is the planetary orbital period.
We use a simple linear regression to fit the linear ephemerides equation to the mid-transit times.
The best solutions obtained were:
\begin{equation}
 T_{\rm min} = 2454375.6251(2) + E \times 1.6284305(2),
\end{equation}
\begin{equation}
 T_{\rm min} = 2455434.3774(7) + E \times 2.423807(3),
\end{equation}
\begin{equation}
 T_{\rm min} = 2455392.3139(9) + E \times 1.430375(2),
\end{equation}
for WASP-5b, WASP-44b, and WASP-46b, respectively. 

\subsection{Transit timing variations}
\label{Transit_timing_variations}

\begin{figure}
\resizebox{\hsize}{!}{\includegraphics[]{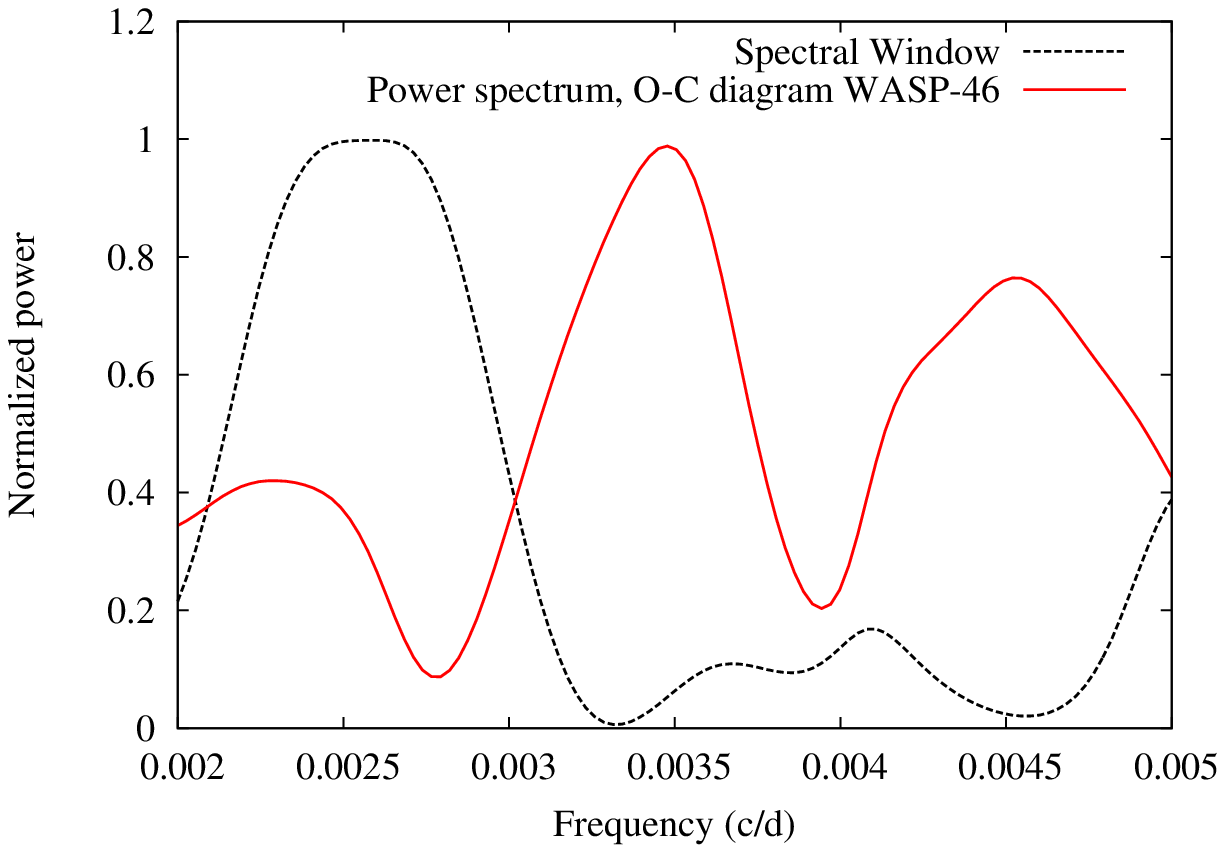}} \\
\caption{The figure shows, in a red continuous line, the periodogram of WASP-46 TTVs zoomed-in around
  the maximum power. On top of this, the spectral window (i.e., the periodicity induced by the sampling
  of the data) is plotted in black dashed lines. For a better comparison, both periodograms have been
  normalized to their respective maximum values.}
\label{wasp-46b_periodogram}
\end{figure}

\begin{figure}
  \resizebox{\hsize}{!}{\includegraphics{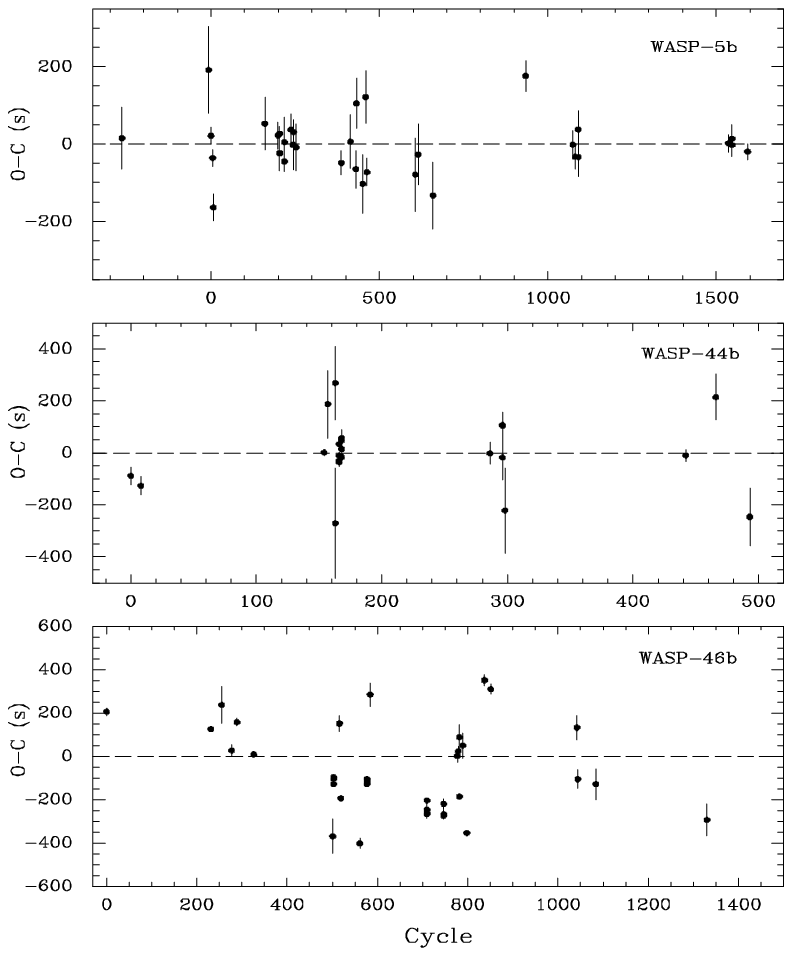}} \\
  \caption{The panels from the top to bottom show the (O-C) diagrams for WASP-5b, WASP-44b,
    and WASP-46b, respectively.}
  \label{oc_diagrams}
\end{figure}

The mid-transit times and the residuals from the fit are shown in Figure~\ref{oc_diagrams} and
listed in Tables ~\ref{wasp5b_timing},~\ref{wasp44b_timing}, and ~\ref{wasp46b_timing}.
In the absence of transit timing variations \citep{Holman2005,Holman2010} we would expect no statistically
significant deviations of the derived [O (\emph{observed}) - C (\emph{calculated})] values from zero. To test
the truthfulness of this we computed $\chi^2_{\rm red}$ for the three systems and found
\mbox{$\chi^2_{\rm red,WASP-5b}$ = 2.1}, \mbox{$\chi^2_{\rm red,WASP-44b}$ = 3.2} and \mbox{$\chi^2_{\rm red,WASP-46b}$ = 61.8}.
Although the three values are sufficiently large to suspect the presence of TTVs, values around
\mbox{$\chi^2_{\rm red}\sim$3} have created already some disagreement between authors (see e.g.,
\citealt{vonEssen2013} but then \citealt{Mislis2015}). Nonetheless, WASP-46b's $\chi^2_{\rm red}$
value is large enough to suspect TTVs. To further investigate this we applied a Lomb-Scargle periodogram
\citep{Lomb1976,Scargle1982,Zechmeister2009} to the (O-C) values. The derived false alarm probabilities (FAP)
are \mbox{FAP$_{\rm WASP-5b}$ = 0.64}, \mbox{FAP$_{\rm WASP-44b}$ = 0.17} and \mbox{FAP$_{\rm WASP-46b}$ = 0.0002}, 
placing again WASP-46b as a good candidate for TTVs. The power spectrum of WASP-46b's periodogram has a peak at
0.0034 $\pm$ 0.0004 c/d (cycles per day), which corresponds to $\sim$295 days. The frequency and error were
computed fitting a Gaussian profile to the maximum peak of the periodogram. As a measurement of the amplitude
of WASP-46's (O-C) diagram we used its standard deviation, with a value of 250.7 seconds. We finally
re-determined the FAP of the TTV signal but using a bootstrap resampling method. For this, we randomly
permuted the mid-transit values along with their errors 5$\times$10$^5$ times, but leaving the epochs fixed.
At each iteration we calculated a Lomb-Scargle periodogram from the permuted (O-C) diagram. We estimated the
FAP as the frequency with which the highest power in the scrambled periodogram exceeds the maximum power in
the original periodogram. Using this more reliable technique, the re-estimated FAP for WASP-46b is 0.04,
inconsistent with the one computed in the usual way but still low. Finally, to characterize WASP-46b's TTVs
we computed the spectral window of its (O-C) diagram. If the signal is associated with the sampling rather than
to the true TTV signal, then the spectral window power spectrum should have a similar peak around 0.0034 c/d.
This was the case. We found one single and isolated peak at 0.0026 $\pm$ 0.0004 c/d, just consistent with the
previously computed TTV signal at the 1-$\sigma$ level (see Figure \ref{wasp-46b_periodogram}). Although
the TTV amplitude is significantly large, we notice that the more distant an (O-C) data point is from zero,
the larger its error bar is. We therefore advise caution to readers who may want to interpret these TTVs.

\begin{table}
\scriptsize
\caption{Mid-transit times for WASP-5b}             
\label{wasp5b_timing}      
\centering                          
\begin{tabular}{l l c r}        
\hline\hline                 
 Cycle  & T(BJD$_{\rm TDB}$)       & (O-C)       & Reference \\    
        & 2450000 $+$              & s           &             \\
\hline                        
-264    & 3945.71962$\pm$0.00093 &    15       &  Gillon et al. 2008 \\
-7      & 4364.2283$\pm$0.0013   &    191      &  Gillon et al. 2008 \\
0       & 4375.62535$\pm$0.00026 &    22       &  Gillon et al. 2008 \\
5       & 4383.7675$\pm$0.0004   &   -36       &  \citet{Anderson2008} \\
7       & 4387.02275$\pm$0.0010  &  -164       &  \citet{Anderson2008} \\
160     & 4636.17459$\pm$0.00079 &    53       & \citet{Fukui2011}       \\
199     & 4699.68303$\pm$0.00040 &    22       & \footnotemark[1]         \\
204     & 4707.82523$\pm$0.00023 &    26       & \citet{Southworth2009b}    \\
204     & 4707.82465$\pm$0.00052 &   -23       & \footnotemark[1]           \\
218     &4730.62243$\pm$0.00031  &   -45       & \citet{Southworth2009b}    \\
218     &4730.62301$\pm$0.00075  &     5       & \footnotemark[1]           \\
237     &4761.56356$\pm$0.00047  &    37       & \footnotemark[1]           \\
244     &4772.96212$\pm$0.00074  &    -2       & \citet{Fukui2011}          \\
245     &4774.59093$\pm$0.00030  &    31       & \footnotemark[1]           \\
253     &4787.61792$\pm$0.00069  &    -9       & \footnotemark[1]           \\
387     &5005.82714$\pm$0.00036  &   -49       & \footnotemark[1]           \\
414     &5049.79540$\pm$0.00080  &     6       & \footnotemark[1]           \\
430     &5075.84947$\pm$0.00056  &   -64       &  \citet{Dragomir2011}      \\
432     &5079.10830$\pm$0.00075  &   105       &  \citet{Fukui2011}         \\
451     &5110.04607$\pm$0.00087  &  -102       &  \citet{Fukui2011}         \\
459     &5123.07611$\pm$0.00079  &   121       &  \citet{Fukui2011}         \\
463     &5129.58759$\pm$0.00042  &   -72       &  \footnotemark[1]          \\
607     &5364.0815$\pm$0.0011    &   -79       &  \citet{Fukui2011}         \\
615     &5377.10955$\pm$0.00091  &   -27       &  \citet{Fukui2011}         \\
659     &5448.75927$\pm$0.0010   &  -133       &  \citet{Dragomir2011}      \\
934     &5896.58124$\pm$0.00046  &   177       &  TRESCA \\
1074    &6124.55945$\pm$0.00042  &    -2       &  TRESCA \\
1082    &6137.58653$\pm$0.00037  &  -33        &  TRESCA \\
1090    &6150.61396$\pm$0.00057  &   -34       &  This study  \\
1090    &6150.61479$\pm$0.00056  &    37       &  This study  \\
1536    &6876.89438$\pm$0.00027  &     1       &  TRESCA \\
1547    &6894.80707$\pm$0.0003415&    -3       &  TRESCA \\
1547    &6894.80726$\pm$0.00042  &    14       &  TRESCA \\
1593    &6969.71467$\pm$0.00024  &   -20       &  TRESCA \\
\hline
\end{tabular}
\footnotemark[1] \citet{Hoyer2012};~
\end{table}

\begin{table}
\scriptsize
\caption{Mid-transit times for WASP-44b}             
\label{wasp44b_timing}      
\centering                          
\begin{tabular}{l l c r}        
 \hline\hline
Cycle  & T(BJD$_{\rm TDB}$)       & (O-C)       & Reference \\    
        & 2450000 $+$              & s           &             \\
\hline                        
 0  &  $5434.37637\pm0.00040$  &  -89     &  \citet{Anderson2012} \\
8   &  $5453.76639\pm0.00042$  & -127     &  \citet{Anderson2012} \\
154 &  $5807.64374\pm0.00013$  &    1     &  \citet{Mancini2013}  \\
157 &  $5814.91731\pm0.00150$  &  187     &  Evans P.(TRESCA)\\
163 &  $5829.45485\pm0.00245$  & -271     &  Lomoz F.(TRESCA) \\
163 &  $5829.46110\pm0.00163$  &  268     &  Lomoz F.(TRESCA)\\
166 &  $5836.72905\pm0.00020$  &  -31     &  \citet{Mancini2013}   \\
166 &  $5836.72979\pm0.00030$  &   33     &  \citet{Mancini2013} \\
166 &  $5836.72900\pm0.00020$  &  -35     &  \citet{Mancini2013} \\
166 &  $5836.72928\pm0.00015$  &   -11    &  \citet{Mancini2013} \\
168 &  $5841.57719\pm0.00035$  &   14     &  \citet{Mancini2013} \\
168 &  $5841.57757\pm0.00046$  &   47     &  \citet{Mancini2013} \\
168 &  $5841.57684\pm0.00028$  &   -16    &  \citet{Mancini2013} \\
168 &  $5841.57769\pm0.00031$  &   57     &  \citet{Mancini2013} \\
286 &  $6127.58626\pm0.00048$  &    -2    &  Sauer T.(TRESCA) \\
296  & $6151.82555\pm0.00060$  &  103     &  This study   \\
296  & $6151.8242\pm0.0010$    &    -18   &  This study   \\
296  & $6151.82559\pm0.00045$  &  106     &  This study   \\
298  & $6156.6694\pm0.0017$    & -222     &  This study   \\
442  & $6505.7001\pm0.0002$    &  -11     &  This study   \\
466  & $6563.87407\pm0.00102$  &   213    &  Evans P.(TRESCA)   \\
493  & $6629.31154\pm0.00127$  &   -247   &  René R.(TRESCA)   \\
\hline
\end{tabular}
\end{table}

\begin{table}
\scriptsize
\caption{Mid-transit times for WASP-46b.}             
\label{wasp46b_timing}      
\centering                          
\begin{tabular}{l l c r}        
\hline\hline                 
Cycle  & T(BJD$_{\rm TDB}$)       & (O-C)       & Reference \\    
        & 2450000 $+$              & s           &             \\
\hline                        
0    &5392.31628 $\pm$ 0.00020 & 206 & \citet{Anderson2012} \\
231  &5722.73197 $\pm$ 0.00013 & 125 & \citet{Ciceri+2016} \\
255  &5757.06225 $\pm$ 0.00098 & 235 & TRESCA \\
277  &5788.52807 $\pm$ 0.00030 &  25 & This work \\
289  &5805.69409 $\pm$ 0.00020 & 157 & \citet{Ciceri+2016}  \\
326  &5858.61624 $\pm$ 0.00011 &   8 & \citet{Ciceri+2016} \\
501  &6108.92750 $\pm$ 0.00091 &-370 & TRESCA \\
503  &6111.79133 $\pm$ 0.00011 &-103 & \citet{Ciceri+2016} \\
503  &6111.79141 $\pm$ 0.00013 & -97 & \citet{Ciceri+2016} \\
503  &6111.79132 $\pm$ 0.00013 &-104 & \citet{Ciceri+2016} \\
503  &6111.79102 $\pm$ 0.00013 &-130 & \citet{Ciceri+2016} \\
516  &6130.38914 $\pm$ 0.00042 & 150 & TRESCA \\
519  &6134.67627 $\pm$ 0.00016 &-195 & TRESCA \\
561  &6194.74962 $\pm$ 0.00027 &-402 & \citet{Ciceri+2016} \\
577  &6217.63904 $\pm$ 0.00013 &-107 & \citet{Ciceri+2016} \\ 
577  &6217.63892 $\pm$ 0.00011 &-117 & \citet{Ciceri+2016} \\ 
577  &6217.63883 $\pm$ 0.00010 &-125 & \citet{Ciceri+2016} \\ 
577  &6217.63877 $\pm$ 0.00012 &-130 & \citet{Ciceri+2016} \\ 
584  &6227.65619 $\pm$ 0.00062 & 284 & TRESCA \\
710  &6407.87778 $\pm$ 0.00014 &-205 & \citet{Ciceri+2016} \\  
710  &6407.87730 $\pm$ 0.00033 &-246 & \citet{Ciceri+2016} \\ 
710  &6407.87711 $\pm$ 0.00017 &-262 & \citet{Ciceri+2016} \\ 
710  &6407.87705 $\pm$ 0.00021 &-268 & \citet{Ciceri+2016} \\
747  &6460.80084 $\pm$ 0.00016 &-275 & \citet{Ciceri+2016} \\  
747  &6460.80090 $\pm$ 0.00020 &-270 & \citet{Ciceri+2016} \\ 
747  &6460.80147 $\pm$ 0.00028 &-221 & \citet{Ciceri+2016} \\ 
747  &6460.80092 $\pm$ 0.00025 &-269 & \citet{Ciceri+2016} \\
777  &6503.71529 $\pm$ 0.00034 &   1 & This work \\
779  &6506.57629 $\pm$ 0.00025 &  23 & This work \\
782  &6510.86498 $\pm$ 0.00013 &-187 & \citet{Ciceri+2016} \\
782  &6510.86816 $\pm$ 0.00067 &  87 & TRESCA \\
789  &6520.88034 $\pm$ 0.00067 &  49 & TRESCA \\
798  &6533.74905 $\pm$ 0.00013 &-354 & \citet{Ciceri+2016} \\
837  &6589.54183 $\pm$ 0.00031 & 350 & TRESCA \\
851  &6609.56661 $\pm$ 0.00027 & 309 & TRESCA \\
1042 &6882.76617 $\pm$ 0.00065 & 131 & TRESCA \\
1044 &6885.62416 $\pm$ 0.00049 &-107 & TRESCA \\
1084 &6942.83890 $\pm$ 0.00083 &-130 & TRESCA \\
1330 &7294.70924 $\pm$ 0.00085 &-295 & TRESCA \\
\hline
\end{tabular}
\end{table}

\subsection{Broadband spectrum}
\label{broadband_spectrum}

\begin{figure}
  \resizebox{\hsize}{!}{\includegraphics{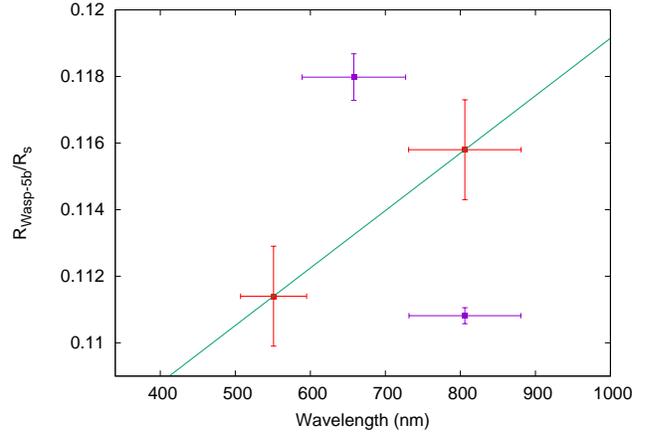}} \\
  \caption{Planet to star radius of WASP-5b as a function of the observed band. The red squares are our
    measurements and the purple squares are the ones available in the literature. The green line
  is a linear fit to our measurements (see text).}
  \label{wasp5b_broadband}
\end{figure}

\begin{figure}
  \resizebox{\hsize}{!}{\includegraphics{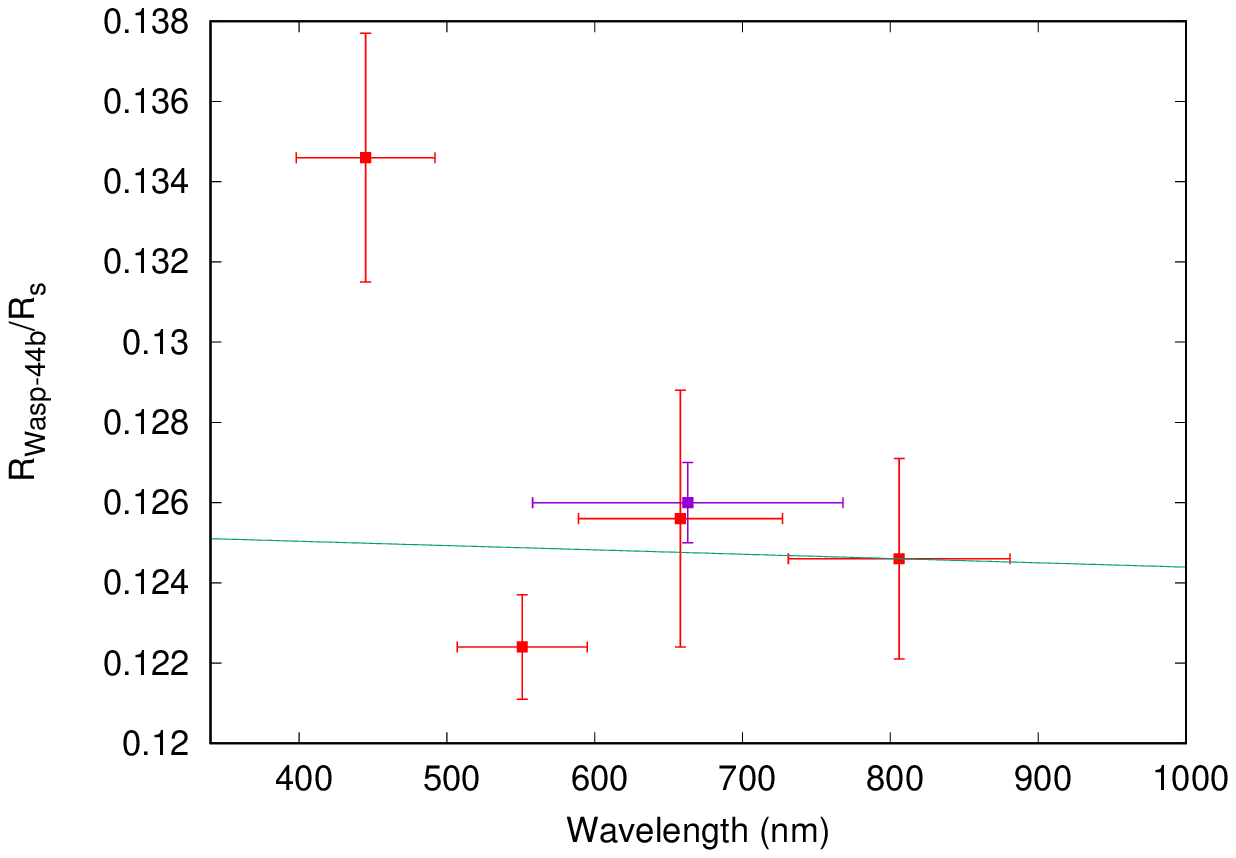}} \\
  \caption{Planet to star radius of WASP-44b as a function of the observed band.
    Plotted as per the description for Figure \ref{wasp5b_broadband}.}
  \label{wasp44b_broadband}
\end{figure}

\begin{figure}
  \resizebox{\hsize}{!}{\includegraphics{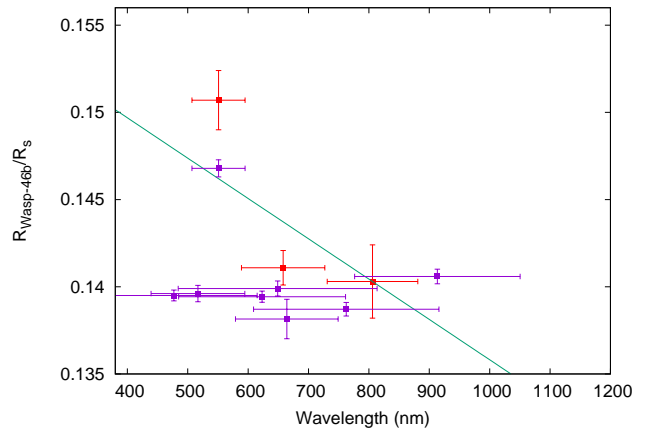}} \\
  \caption{Planet to star radius of WASP-46b as a function of the observed band.
    Plotted as per the description for Figure \ref{wasp5b_broadband}.}
  \label{wasp46b_broadband}
\end{figure}

Multi-band transit observations allow to study the fractional radii variation as a function of the
wavelength. The planetary radius $R_p$ derived from transit observations may vary with the wavelength,
as it can appear slightly larger when observed at wavelengths where the atmosphere contains strong
opacity sources \citep{Burrows2000, SeagerSasselov2000}. These changes are at first approximation the
broad-band transmission spectrum \citep[e.g.][]{Nikolov2013}. We fitted our light curves again,
  but now we introduced as priors the calculated common parameters shown in Table~\ref{weighted_mean_parameters}
  and the orbital periods. This improves the quality of the fractional radius determination. The improved results
  for this parameter are shown in Tables ~\ref{wasp5b_table_results}, \ref{wasp44b_table_results}, and
  \ref{wasp46b_table_results} for WASP-5b, WASP-44b, and WASP-46b, respectively. Figures ~\ref{wasp5b_broadband},
\ref{wasp44b_broadband}, and \ref{wasp46b_broadband} show the resulting fractional radii variation as a
function of wavelength using our measurements (red squares) and the ones available in the literature (pink squares)
for the planets WASP-5b, WASP-44b, and WASP-46b, respectively. The horizontal error bars represent the FWHM of the used
filters. The superimposed dotted green line in these figures is the linear fit using only our measurements
constraining the fit to cross our measurement with the longest wavelength. We calculated the slopes of
these lines to investigate some trends in our measurements. The values of the slopes are listed in
Table~\ref{LOG_observations}.

In the case of WASP-5b, there are not enough measurements to have an unambiguous view to describe a
conclusive scenario for any correlation between the planetary radius and wavelength \citep[e.g.][]{Jha2000}.
In the case of WASP-44b, our values are consistent with a flat spectrum similar to that of \citet{Turner2016}
(see their figure 9). Our calculated slope (see Table~\ref{LOG_observations}) is consistent with zero.
A flat spectrum is indicative of the presence of clouds in the upper atmosphere which prevent observation of any
spectral features \citep{Turner2016, SeagerSasselov2000}. In the case of WASP-46b, we see indications of an
increase of radius ($4\sigma$) at the shortest wavelengths, a hint of possible Rayleigh scattering in
the atmosphere of this planet \citep{Lecavelier2008}. We notice that \citet{Ciceri+2016} found no strong
evidence of Rayleight scattering in the atmosphere of WASP-46b (see their figure 8). They found a slope
with maximum inclination of $ m = -1.17 \times 10^{-5}$ for WASP-46b ($1/2$ of our slope, see
Table~\ref{LOG_observations}). Although there is rotational modulation on this system \citep{Anderson2012},
the stellar activity is not sufficient to explain this difference, so further measurements are necessary
confirm or rule out this scenario. We also notice that the V-band light curve of WASP-46b is missing the flat
part after egress (see Figure \ref{wasp-46b_lc}), thus preventing from strongly constraining the baseline value,
although our measurement is consistent within $2 \sigma$ with that of \citet{Anderson2012}. As pointed out by
\citet{Gibson2014}, choosing an incorrect noise model might artificially lower the uncertainties which
is equivalent to getting the normalization of the light curve wrong, thus such high slope could be in part
explained by the lack of observations after egress. We therefore advise caution to readers who may want to
interpret this significant variation of the broadband spectrum of WASP-46b. Although we recognize the
  V-band planetary to star radii ratio of WASP-46b is significantly larger than in the other bands, additional
  measurements will be needed to confirm this finding due to our lack of observations after egress.\\

  As an alternative way of quantifying how much do the planetary to star radii ratio and wavelength correlate
  to each other, we made use of the Pearson's correlation coefficient, $r$. The derived values are as follows:
  $r_{WASP-44b}$ = -0.61, $r_{WASP-46b}$ = 0.11, and  $r_{WASP-5b}$ = 0.05. These coefficients indicate that there
  is no correlation between the wavelength and planetary to star radii ratio for the systems WASP-5b and
  WASP-46b (values are close to zero). The coefficient $r_{WASP-44b}$ indicates that there is a significant
  anti-correlation for WASP-44b. In any case, we consider it unlikely that the amplitude of the variability
  in WASP-44b could vary with wavelength by more than 10 scale heights, a sensible limit for expected variability
  when Rayleigh scattering is being investigated \citep{Sing2016}.

\section{Summary}
\label{summary}

In this paper we have presented multi-band photometry of three hot Jupiters: WASP-5b, WASP-44b, and WASP-46b.
The data were collected as part of an observational program to characterize hot Jupiters which has been carried
out with the facilities of the Pico dos Dias Observatory in Brazil. We performed a detailed analysis of the
planetary transits using the following programs: EXOFAST \citep{Eastman2013}, TAP \citep{Gazak2012}, and,
JKTEBOP \citep{Southworth2008}. Using observations in the $V-$, $B-$, $R_{C}-$, and $I_{C}-$bands we were able
to improve the geometrical and physical parameters of WASP-5b and WASP-46b (see Tables~
\ref{wasp5b_table_results} and \ref{wasp46b_table_results}), in particular the ratio between the semi-major
axis and the stellar radius for WASP-46b $a/R_* = 5.76\pm0.09$. The parameters we derived are consistent
with previous studies (see Table~\ref{weighted_mean_parameters})

We obtained improved linear ephemerides for WASP-5b, WASP-44b, and WASP-46b using our mid-transit times
measurements together with all the other measurements available in the literature
(see Section~\ref{Transit_ephemerides}). The analysis of the residuals from the linear fit, the (O-C) diagram,
doesn't show a clear indication of transit timing variation for WASP-5b and WASP-44b (see Figure \ref{oc_diagrams}).
The (O-C) diagram of WASP-46b shows variation with a semi-amplitude of $\sim$10 min. We proved
this variation is most likely due to the sampling of the observations, but further additional transit
time measurements of this system are necessary to secure this finding.

Finally, we studied the variation with wavelength of the planet to star radius for WASP-5b, WASP-44b, and
WASP-46b. We don't have enough measurements to describe the broad-band spectrum of WASP-5b. In the
case of WASP-44b, our values are consistent with a flat spectrum similar to that of previous studies. In the case of
WASP-46b, we found marginal evidence of Rayleigh scattering which is not in agreement with a previous study.
Further measurements are necessary to confirm or rule out this finding.

\section*{ACKNOWLEDGEMENTS}
M. Moyano thanks Joseph Carson and an anonymous referee for some helpful suggestions and for a critical
reading of the original version of the paper. L. A. Almeida acknowledges support from CAPES and Funda\c{c}\~ao
de Amparo \`a Pesquisa do Estado de S\~{a}o Paulo - FAPESP (2013/18245-0 and 2012/09716-6). C. von Essen
acknowledges funding for the Stellar Astrophysics Centre, provided by The Danish National Research Foundation
(Grant DNRF106).

\label{lastpage}

\begin{thebibliography}{}
  
\bibitem[\protect\citeauthoryear{Anderson et al.}{2008}]{Anderson2008}
Anderson D.~R., et al., 2008, MNRAS, 387, L4
\bibitem[\protect\citeauthoryear{Anderson et al.}{2012}]{Anderson2012}
Anderson D.~R., et al., 2012, MNRAS, 422, 1988
\bibitem[\protect\citeauthoryear{Burrows et al.}{2000}]{Burrows2000} 
Burrows A. et al., 2000, ApJ, 534, L97
 \bibitem[\protect\citeauthoryear{Carter \& Winn}{2009}]{Carter2009}
Carter J.~A., Winn J.~N., 2009, ApJ, 704, 51
\bibitem[\protect\citeauthoryear{Charbonneau et al.}{2000}]{Charbonneau2000}
Charbonneau D., Brown T.~M., Latham D.~W., Mayor M., 2000, ApJ, 529, L45
\bibitem[\protect\citeauthoryear{Ciceri et al.}{2016}]{Ciceri+2016} 
Ciceri S., et al., 2016, MNRAS, 456, 990 
\bibitem[\protect\citeauthoryear{Claret \& Bloemen}{2011}]{Claret2011} 
Claret A., Bloemen S., 2011, A\&A, 529, A75 
\bibitem[\protect\citeauthoryear{Dragomir et al.}{2011}]{Dragomir2011} 
Dragomir D., et al., 2011, AJ, 142, 115 
\bibitem[\protect\citeauthoryear{Eastman, Siverd, \& Gaudi}{2010}]{Eastman2010}
Eastman J., Siverd R., Gaudi B.~S., 2010, PASP, 122, 935
\bibitem[\protect\citeauthoryear{Eastman, Gaudi, \& Agol}{2013}]{Eastman2013}
  Eastman J., Gaudi B.~S., Agol E., 2013, PASP, 125, 83
\bibitem[\protect\citeauthoryear{Espinoza \& Jord{\'a}n}{2015}]{Espinoza2015}
  Espinoza N., Jord{\'a}n A., 2015, MNRAS, 450, 1879 
\bibitem[\protect\citeauthoryear{Ford}{2005}]{Ford2005}
Ford E.~B., 2005, AJ, 129, 1706
\bibitem[\protect\citeauthoryear{Ford}{2006}]{Ford2006}
Ford E.~B., 2006, ApJ, 642, 505
\bibitem[\protect\citeauthoryear{Ford \& Rasio}{2008}]{FordRasio2008}
Ford, E.~B.,  Rasio, F.~A., 2008, ApJ, 686, 621
\bibitem[\protect\citeauthoryear{Fukui et al.}{2011}]{Fukui2011} 
Fukui A., et al., 2011, PASJ, 63, 287
\bibitem[\protect\citeauthoryear{Gazak et al.}{2012}]{Gazak2012}
Gazak J.~Z., Johnson J.~A., Tonry J., Dragomir D., Eastman J., Mann A.~W.,
Agol E., 2012, AdAst, 2012,
\bibitem[\protect\citeauthoryear{Gibson}{2014}]{Gibson2014}
  Gibson N.~P., 2014, MNRAS, 445, 3401
\bibitem[\protect\citeauthoryear{Holman \& Murray}{2005}]{Holman2005} 
Holman M.~J., Murray N.~W., 2005, Sci, 307, 1288 
\bibitem[\protect\citeauthoryear{Holman et al.}{2010}]{Holman2010} 
Holman M.~J., et al., 2010, Sci, 330, 51 
\bibitem[\protect\citeauthoryear{Hoyer, Rojo, \& L{\'o}pez-Morales}{2012}]{Hoyer2012} 
Hoyer S., Rojo P., L{\'o}pez-Morales M., 2012, ApJ, 748, 22 
\bibitem[\protect\citeauthoryear{Jha et al.}{2000}]{Jha2000}
Jha S. et al., 2000, ApJ, 540, L45
\bibitem[\protect\citeauthoryear{Jenkins, Caldwell, \& Borucki}{2002}]{Jenkins2002}
  Jenkins J.~M., Caldwell D.~A., Borucki W.~J., 2002, ApJ, 564, 495
\bibitem[\protect\citeauthoryear{Lecavelier Des Etangs et al.}{2008}]{Lecavelier2008}
  Lecavelier Des Etangs A., Vidal-Madjar A., D{\'e}sert J.-M., Sing D., 2008, A\&A, 485, 865 
\bibitem[\protect\citeauthoryear{Lomb}{1976}]{Lomb1976}
Lomb N.~R., 1976, Ap\&SS, 39, 447
  \bibitem[\protect\citeauthoryear{Mancini et al.}{2013}]{Mancini2013} 
Mancini L., et al., 2013, MNRAS, 430, 2932 
\bibitem[\protect\citeauthoryear{Mandel \& Agol}{2002}]{Mandel2002}
Mandel K., Agol E., 2002, ApJ, 580, L171
\bibitem[\protect\citeauthoryear{Mayor \& Queloz}{1995}]{Mayor1995}
Mayor M., Queloz D., 1995, Nature, 378, 355
\bibitem[\protect\citeauthoryear{Mislis et al.}{2015}]{Mislis2015}
Mislis D., et al., 2015, MNRAS, 448, 2617 
\bibitem[\protect\citeauthoryear{Nikolov et al.}{2013}]{Nikolov2013}
Nikolov N., Chen G., Fortney J.~J., Mancini L., Southworth J., van Boekel R., Henning T., 2013, A\&A, 553, A26 
\bibitem[\protect\citeauthoryear{Pont, Zucker, \& Queloz}{2006}]{Pont2006}
Pont F., Zucker S., Queloz D., 2006, MNRAS, 373, 231
\bibitem[\protect\citeauthoryear{Scargle}{1982}]{Scargle1982}
Scargle J.~D., 1982, ApJ, 263, 835 
\bibitem[\protect\citeauthoryear{Schneider et al.}{2011}]{Schneider2011}
Schneider J., Dedieu C., Le Sidaner P., Savalle R., Zolotukhin I., 2011, A\&A, 532, A79
\bibitem[\protect\citeauthoryear{Seager \& Sasselov}{2000}]{SeagerSasselov2000}
Seager S., Sasselov D. D., 2000, ApJ, 537, 916
\bibitem[\protect\citeauthoryear{Seager \& Mall{\'e}n-Ornelas}{2003}]{Seager2003}
Seager S., Mall{\'e}n-Ornelas G., 2003, ApJ, 585, 1038
\bibitem[\protect\citeauthoryear{Seager \& Deming}{2010}]{Seager2010ARAA} 
Seager S., Deming D., 2010, ARA\&A, 48, 631 
\bibitem[Sing et al.(2016)]{Sing2016} 
  Sing, D.~K., Fortney, J.~J., Nikolov, N., et al. 2016, Nature, 529, 59
\bibitem[\protect\citeauthoryear{Southworth}{2008}]{Southworth2008}
Southworth J., 2008, MNRAS, 386, 1644
\bibitem[\protect\citeauthoryear{Southworth}{2009}]{Southworth2009}
Southworth J., 2009, MNRAS, 394, 272
\bibitem[\protect\citeauthoryear{Southworth et al.}{2009}]{Southworth2009b} 
Southworth J., et al., 2009, MNRAS, 396, 1023 
\bibitem[\protect\citeauthoryear{Stetson}{1987}]{Stetson1987}
Stetson P.~B., 1987, PASP, 99, 191
\bibitem[\protect\citeauthoryear{Swain et al.}{2010}]{Swain2010}
Swain M.~R., et al., 2010, Natur, 463, 637
\bibitem[\protect\citeauthoryear{Tegmark et al.}{2004}]{Tegmark2004}
  Tegmark M., et al., 2004, PhRvD, 69, 103501 
\bibitem[\protect\citeauthoryear{Ter Braak}{2006}]{terBraak2006}
ter Braak, C. J. F. 2006, Statistics and Computing, 16, 239
\bibitem[\protect\citeauthoryear{Thies et al.}{2011}]{Thies2011}
  Thies, I.,et al., 2011, MNRAS, 417(3), 1817
\bibitem[\protect\citeauthoryear{Triaud et al.}{2010}]{Triaud2010}
  Triaud A.~H.~M.~J., et al., 2010, A\&A, 524, A25
\bibitem[\protect\citeauthoryear{Turner et al.}{2016}]{Turner2016}
  Turner J.~D., et al., 2016, MNRAS, 459, 789
\bibitem[\protect\citeauthoryear{Tutukov \& Fedorova}{2012}]{TutukovFedorova2012}
Tutukov, A. V., Fedorova, A.~V., 2012, Astronomy Reports, 56, 305
\bibitem[\protect\citeauthoryear{Valsecchi \& Rasio}{2014}]{ValsecchiRasio2014} 
Valsecchi, F., Rasio, F.~A., 2014, ApJ 786, 102
\bibitem[\protect\citeauthoryear{von Essen et al.}{2013}]{vonEssen2013}
von Essen C., Schr{\"o}ter S., Agol E., Schmitt J.~H.~M.~M., 2013, A\&A, 555, A92
\bibitem[\protect\citeauthoryear{Wolszczan \& Frail}{1992}]{WolszczanFrail1992}
Wolszczan, A., Frail D.~A., 1992, Nature, 355, 145
\bibitem[\protect\citeauthoryear{Wolszczan}{1994}]{Wolszczan1994}
Wolszczan, A., 1994, Science, 264, 538
\bibitem[\protect\citeauthoryear{Zechmeister \& K{\"u}rster}{2009}]{Zechmeister2009}
Zechmeister M., K{\"u}rster M., 2009, A\&A, 496, 577
\end{thebibliography}
\end{document}